# Raman Spectroscopy of Nanostructures and Nanosized Materials

Gwénaël Gouadec and Philippe Colomban*
Groupe Nanophases & Solides Hétérogènes
Laboratoire de Dynamique, Interaction et Réactivité, UMR7075 CNRS – Université Pierre et Marie Curie-Paris 6, 2 rue Henry Dunant, 94320 Thiais, France

Fax 33 1 49 78 11 18
*corresponding author
Philippe.colomban@glvt-cnrs.fr

**Abstract**
The interest of micro and tip-enhanced Raman spectroscopy to analyze nanosized and nanostructured materials, chiefly semiconductors, oxides and pristine or functionalized carbon nanotubes, is reviewed at the light of the contributions to this special issue. Particular attention is paid to the fact that chemical reactions, size or shape distribution, defects, strain and couplings may add to nano-dimensionality in defining the Raman signature.

Raman spectroscopy status as a nanomaterials investigation technique is presented at the light of the contributions to this special issue.

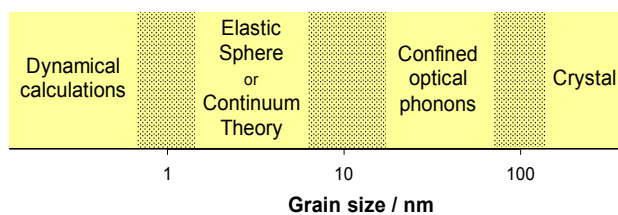

**Raman Spectroscopy of Nanostructures and Nanosized Materials**

Gwénaël Gouadec and Philippe Colomban

## INTRODUCTION

There is no saying how many promises (but also concerns) were brought by with the emergency of nanomaterials in the last few years. This is illustrated in Figure 1, based on a search carried on the "Web of Science" bibliographic portal. Over the period 1995 to 2006, the number of Research papers dedicated to a "nano" thematic thus increased from about 3000 per year to almost 40 000, with a marked acceleration since year 2000. In the same period, the proportion of these articles producing Raman spectroscopy results steadily increased from one out of 25 to one out of 20. Solid state physicists working on semi-conductors and carbon were the first to systematize Raman spectroscopy use to investigate nanosized samples. They were soon followed by oxides specialists and chemists interested with reactivity and surface phenomena. Thus, not only is Raman Spectroscopy a choice method to investigate nanomaterials but it has even been increasing its influence in this regard[1].

*Why such interest in the Raman analysis of nanomaterials ?*

Even though conventional Raman micro-spectrometers cannot provide laser spots much smaller than one micron in diameter, one has to remember that the spectra actually stem from the bonds vibrations. This intrinsic nano-probing makes Raman spectroscopy very sensitive to the short-range structure - including in amorphous/glassy materials[1-3] - and offers a "bottom-up" approach of nanostructured materials that comes as a good complement to methods like TEM or X-ray diffraction.

Because almost no sample preparation is needed, researchers with an easy access to a spectrometer might find it worth the try to record a Raman spectrum of their nanomaterials. This could provide them with a phase identification (reference peaks are disturbed but often recognizable after a size reduction) and, possibly, a size estimation. The latter would be based on either Richter, Wang and Ley's description of confined optical modes (the Phonon Confinement Model: PCM[4]) or Lamb's description of the low frequency oscillations in free elastic spheres (the Elastic Sphere Model: ESM[5-7]).

Researchers from spectroscopy-dedicated laboratories will seek a deeper interpretation of the spectra, taking the following facts into account[1]:

- The very high volume fraction of interfaces (from a practical point of view, at least one atomic layer) and grain boundary interphases in nanomaterials (Figure **2**) makes it impossible to discard their specific contributions.

- Over time, the PCM and ESM have been improved to account for particles non sphericality (imperfect spheres, nanowires, etc), size distribution or anisotropy but they often remain too approximate to perfectly fit experimental spectra. Besides, neither model considers the possibility of close particles to interact, nor that for non "free-standing" nanoparticles to couple with the matrix.

- The ESM and PCM rely on hypotheses that are unsatisfied for dimensions smaller than a certain limit (Figure **3**):
  **i)** the PCM assumes that phonons (collective modes) propagate in nanoparticles with a wavevector uncertainty (a consequence of Heisenberg's principle) that can be accounted for with a weighed exploration of "bulk" dispersion curves. By nature, phonons are however a consequence of long distance periodicity and they must gradually disappear as size goes decreasing. Only macroscopic (continuum) theories applied with appropriate mechanical and electrostatic grain boundary conditions[8,9] might then describe vibrations in the optical range.

**ii)** the ESM uses Elastodynamics theory and the underlying matter homogeneity condition only applies to particles much larger than the interatomic distances. Molecular-dynamics simulations [10,11] (currently limited to 5 nm grain size by calculation power[12]) are necessary to model the spectra of smaller particles.

Obviously, the *JRS* did not wait until today before publishing on nanostructured materials (more than 30 articles in year 2006) but it is the first time it proposes a special issue on this high interest subject (for general information on nanomaterials, see for instance references [13-16]). Its collection of 19 original and review papers, with contributors from 16 different countries, covers most of the research area, with the exception of embedded metal particles[17,18], the much debated boson peak characterizing amorphous materials at the nanometer scale[19,20] and Surface-Enhanced Raman Spectroscopy (SERS), which is briefly mentioned in Ref. [21] but was the subject of a previous special issue[22]. The acknowledged ambition is to participate in understanding how Raman spectroscopy can help better design and characterize nanomaterials.

**REVIEW PAPERS**

We open this special issue with Review papers that should altogether provide the interested reader with a general knowledge on how nanomaterials spectra are currently dealt with by the scientific community:

- Arora *et al.* (*Indira Ghandi Centre for Atomic Resarch*)[23] review on the spatial confinement of optical phonons (including in resonant conditions). A special attention is paid to the Gaussian distribution coefficient used in the popular PCM to account for the Raman shift and asymmetric broadening observed in nanocrystalline materials. The discussion is illustrated with superlattices ($GaAs/Al_xGa_{1-x}As$), nanowires (Si, Ge, ZnO, $SnO_2$, etc), powdered or film-deposited oxide and semiconductor nanoparticles (ZnO, $ThO_2$, $TiO_2$, $CeO_2$, Si, GaAs, ZnSe, graphite) and glass-embedded quantum dots.
- Rolo and Vasilevskiy (*Universidad do Minho*)[12] also focus on optical phonon confinement (in cubic polar semiconductors) but in the framework of macroscopic continuum models. Their review first exposes how simultaneously solving the continuous equation of motion and Poisson's equation leads to confined phonons with a mixed longitudinal/transversal/interfacial nature. They proceed with the description of confined excitons and the effect their interaction with confined phonons has on resonant and non-resonant Raman spectra. The theoretical predictions are confronted with experimental results on small sized (~3-6 nm) nanocrystals from the II-VI and III-V families.
- G. Irmer (*Teschnische Universität Bergakademie Freiberg*)[24] completes the description of polar semiconductor nanocrystals. He reviews on the confined optical modes (notably their use for Raman-based grain-size determination), surface (Fröhlich) modes and their interactions with surface plasmons. An "effective medium" model of embedded nanoparticles in interaction is then adapted to account for the experimental signature of GaP, InP and GaAs samples with either columnar or parallel nanoporosity.
- Ivanda *et al.* (*Ruder Boskovic Institute, University of Zagreb, Universita di Trento, CNR & Universität Würzburg*)[25] shift attention onto the low wavenumber domain in nanoparticles Raman spectra. They first recall how Lamb's 1882 theoretical description of a free elastic sphere oscillations, the so-called spheroid and torsional modes, was later transposed to nanoparticles acoustic modes so that they could be used for grain size determination. Taking into account the dependence of Raman cross section on crystallite size, they elaborate on how spectra from polydisperse nanoparticles can be fitted to return their size distribution

- and proceed with case studies on either free or embedded semiconductor ($CdS_xSe_{1-x}$) and oxide ($SnO_2$, $TiO_2$, $HfO_2$) nanoparticles. Ivanda *et al.* conclude that Raman spectroscopy makes it theoretically possible to measure longitudinal and transverse sound velocities in nanoparticles of known diameter.
- Boolchand *et al.* (*University of Cincinnati*)[26] review their achievements, shared with many past and present co-workers, in using Raman spectroscopy to characterize intermediate phases in chalcogenide and oxide glass systems (these phases correspond to stress-free networks that self-organize in disordered systems according to the local or medium range structure; notably the coordination number).
- R. Graupner's review (*Universität Erlangen*)[27] is dedicated to the characterization of covalently-functionalized metallic and semiconducting Single-Wall Carbon Nanotubes (SWCNTs). Raman spectroscopy is a lead investigation techniques for CNTs because a double resonance phenomenon not only allows for the analysis of a single tube (due to signal enhancement) but also, as it involves electronic transitions, gives a direct insight into the electronic properties[28]. However, CNTs need to be functionalized for handling (to prevent agglomeration), sorting (according to electronic or structural characteristics) and property adjustments. The biggest challenge from the Raman spectroscopy point of view then becomes to discriminate signal modifications due to functionalizing groups from those indicating structural modifications of the CNTs.
- Hayazawa *et al.* (*RIKEN, Osaka University, Japan Corporation of Science & Technology*)[21] present recent progress in Tip-Enhanced Raman Spectroscopy (TERS), their practical interest in the technique being the possibility to detect strain fluctuations in silicon layers (such strains are very important to control because they strongly enhance the carriers mobility). TERS originally came as an implementation of Raman Near-Field Scanning Optical Microscopy (RNSOM), where the optical field was confined to a small aperture at the tip of a metal-coated optical fibre brought extremely close to the sample. This allowed to overcome resolution limiting diffraction phenomena (the $\lambda/2$ limit of Abbé criterion) but only a faint signal could be collected due to the fibre cut-off [29,30]. In TERS, the optical fibre has been replaced with an apertureless metallic tip, which favors a surface enhancement of the Raman signal (the so-called SERS effect).
- Cao *et al.* (*Drexel University*)[31] sign the last of the eight review articles with a contribution on the Raman enhancement provided by semiconductor nanowires and nanocones. This enhancement is diameter, wavelength and polarization dependant and models for its interpretation are proposed, which could be used for engineered photonic and sensing applications.

**RESEARCH PAPERS**

*Carbon Nanotubes*

The Raman spectrum of single-wall CNTs is now well-understood (see the abovementioned review from R. Graupner[27] and next four papers introductions[32-35]) but work is in progress to interpret those from multi-walled (MWCNTs) and functionalized/adsorbing nanotubes.
- Kuzmany *et al.* (*Universität Wien, Budapest University & National Hellenic Research Foundation*)[32] demonstrate with the examples of anthracene, $C_{60}$ and $C_{59}$-N heterofullerene molecules encapsulated in SWCNTs that Raman spectra reveal the tubes structural modifications and can be used to estimate the filling rate. Information of this nature is of great value as their hollow inside should make SWCNTs perfect templates for the growth of other nanomaterials.

- Puech *et al.* (*CNRS-Université Paul Sabatier*)[33] address the lineshape modifications observed on the Raman spectra of MWCNTs exposed to air and methanol atmospheres or to electronic irradiation. They focus on $G'_{2D}$ second order mode and find that the inside and outside walls of MWCNTs give distinct Raman contributions.
- M. Amer (*Wright State University*)[34] confronts semi-empirical quantum simulations of water/methanol absorption on CNTs/fullerenes with experimental Raman spectra. The absorption is shown not only to depend on the absorbate but also on entropy factors (concentration/pressure of the absorbate in the liquid mixture containing either CNTs or fullerenes). These results are important in view of applications using CNTs and fullerenes in selectively absorbing nano-filters and substrates.
- Osswald *et al.* (*Drexel University*)[35] use Raman spectroscopy to compare the efficiency of an acidic attack and a flash oxidation in creating defect sites on the surface of MWCNTs (defect sites are necessary for later functionalization). They show that the two routes have quite similar yields but that flash oxidation causes much less damage to the tubes structure.

*Oxides*

- Korotcov *et al.* (*National Taiwan Universities*)[36] introduce stress-dependent dispersion curves in the PCM to fit the Raman spectra of well-aligned $RuO_2$ and $IrO_2$ whiskers (~50x50 to 100x100nm$^2$) grown on different substrates. They correlate the derived stress with theoretical calculations to get an indirect determination of the preferential growth direction.
- Popovic *et al.* (*Belgrade Institute of Physics and SCK CEN*)[37] propose a modified PCM introducing size distribution, stress, non stoichiometry and anharmonicity contributions into the usual grain-size-parameterized expression for Raman intensity. They apply it to the Raman signatures of doped ceria (a model material for the study of size effects) nanopowder and silicon nano-wires.
- Pagnier *et al.* (*ENSEEG*)[38] illustrate with different nano-oxides how Raman scattering turns out to be a convenient surface probing tool which, contrarily to alternative surface analysis techniques, does not require ultra-vacuum conditions. They demonstrate the grain size effect on the phase transition and surface reactivity of $ZrO_2$, $WO_3$ and $SnO_2$ oxides.
- Kesavamoorthy & Sivasubramanian (*Indira Ghandi Center for Atomic Research*)[39] discuss the Raman and luminescence spectra of a PMN piezoelectric perovskite. The former are used to characterize a core-shell nano-structuring, assuming that the two low frequency modes detected are the surface spheroid mode of either the core or the shell. Note that the investigated polycrystalline sample and the intermediate phases presented above[26] are completely opposite among nanostructured materials: here, non-stoichiometry introduces nanodisordering in a reference crystal whereas intermediate phases reflect a nanostructuration of amorphous systems.

*Spectra modeling in nanomaterials*

- Murray *et al.* (*University of British Columbia Okanagan & CNRS-Université de Bourgogne*)[40] propose three different numerical calculations for the polarizability of vibrating nanoparticles. Their results show reasonable agreement and pave the way to the modeling of the low wavenumber Raman signal of metal and dielectric nanoparticles embedded in transparent matrices.
- Gu *et al.* (*Nanyang Technological University & East China Normal University*)[41] offer a "broken bond rule" to account for the optical redshift in nanostructures. Starting with the vibrational energy of a single bond, which they modify according to the bond order reduction (and subsequent bond contraction) in the three atomic layers in contact with a surface (or a defect), they end up with an expression for Raman shifts as a function of the grain size and temperature. This model does not account for the lineshape but still comes as an alternative to the popular PCM.

*High spatial resolution Raman spectroscopy*

- Lee *et al.* (*The University of Akron, Wright Patterson AFB & NIST*)[42] report on the Tip-Enhanced Raman Spectroscopy (TERS) of strained silicon (the same topic as in the review from Hayazawa *et al.*[21]). Raman mapping at high resolution (~20 nm) was achieved thanks to a careful optimization of the beam polarization (to get the best near field to far field contrast possible).

## CONCLUSION AND PERSPECTIVES

Raman spectroscopy has undoubtedly imposed as a nanomaterial probing solution. On the one hand, a phase identification is often readily available from the position of the different peaks. On the other hand, and despite their apparent simplicity, the popular PCM and ESM models may provide great value information on typical dimensions. Thorough interpretation of the spectral features have even been proposed for single wall carbon nanotubes and, to a lower extent, nanostructured semiconductors (quantum dots, wells and heterostructures).

There is, however, much work ahead before spectra are fully understood. First of all, any specific application requires some engineering, for instance a functionalization in the case of CNTs, and the resulting modifications of the Raman spectrum must be carefully interpreted. Moreover, most nanostructured materials are a complex assembly of moieties with distributed characteristics (structure, size, residual stress, etc) and that couple with each other or with the matrix they are embedded in. Separating the impact of each parameter on Raman signal is - and will remain - far from trivial. All the more that light penetration, hence the probed volume, is often indeterminate in the lack of absorption coefficients.

Increasing the spatial resolution with Tip-Enhanced Raman Spectroscopy is a promising orientation but a great challenge as far as microspectroscopy is concerned probably is to go further in combining the physical and chemical approaches to the problem: Raman scattering process obviously is of physical nature but nanostructures and nanosized materials usually have very complex chemistry, their moieties being subject to size-governed phase transitions and having extremely high surface reactivity. The interest of controlling these moieties with "smart" Raman maps, showing the variation of a property - calculated after input of fitted Raman parameters in the appropriate model - as a function of the position in a sample has been demonstrated for SiC fibres and composites[43,44]. It can be predicted that such studies will become all the more common to characterize the chemical and physical heterogeneity of processed or aged nanomaterials that "good" models will be made available by the material science community.

## ACKNOWLEGMENTS
The authors would like to thank all contributors and reviewers for their enthusiasm and the good care they took at this collective work completion.

FIGURE CAPTIONS

**Fig. 1 :** Number of articles found to match various topic combinations searching the "Web of Science" portal (Science Citation Index expanded ; English articles only).

**Fig. 2 :** The "outer shell to bulk" volume ratio in nanospheres with diameter **D** and an external interphase of constant thickness **t**. The schematic details the different contributions to the Raman spectrum.

**Fig. 3 :** The descriptions of vibrations in spherical (modified versions exist for other shapes) nanocrystals[1]. The actual applicability ranges depend on the material and dotted areas indicate possible overlapping. ESM: Elastic Sphere Modeling of low wavenumber peaks; PCM: Phonon Confinement Model for optical modes.

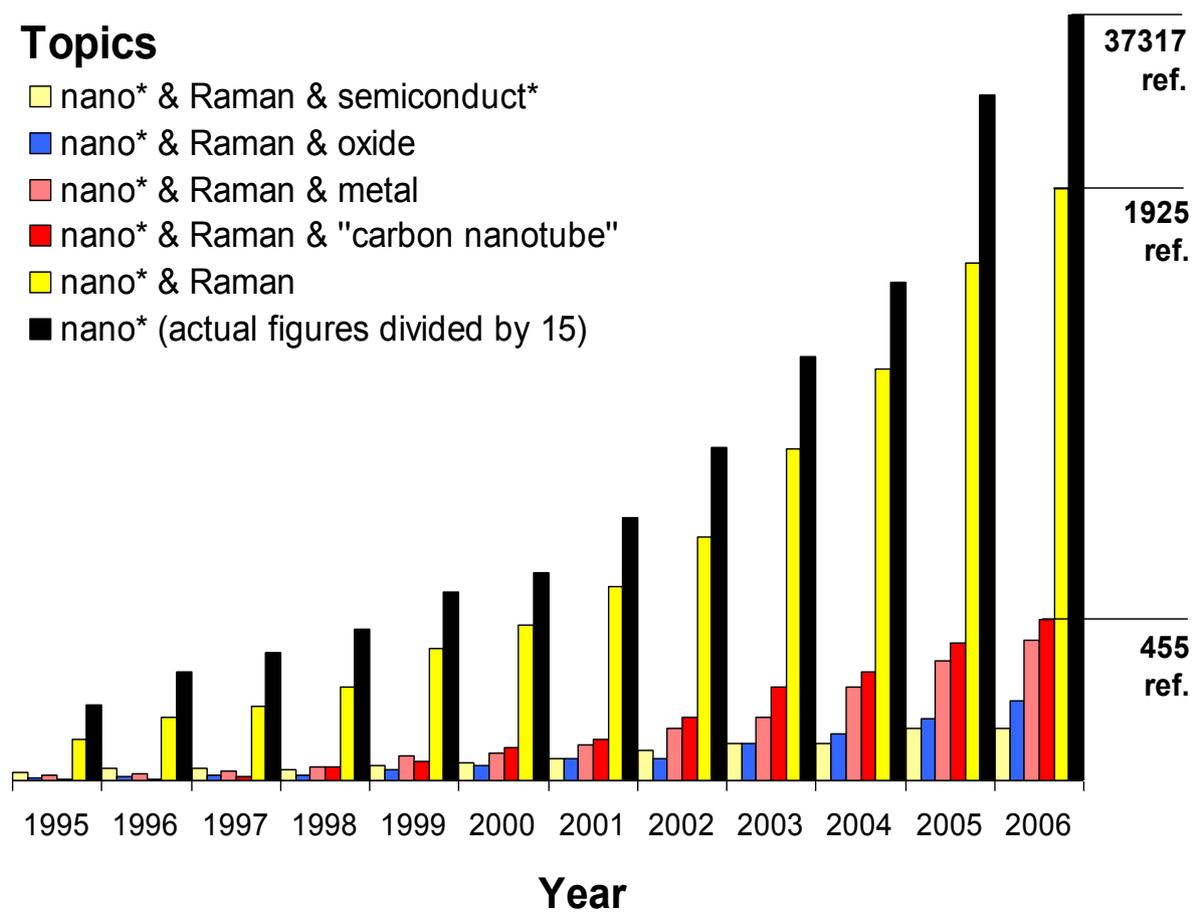

Fig. 1

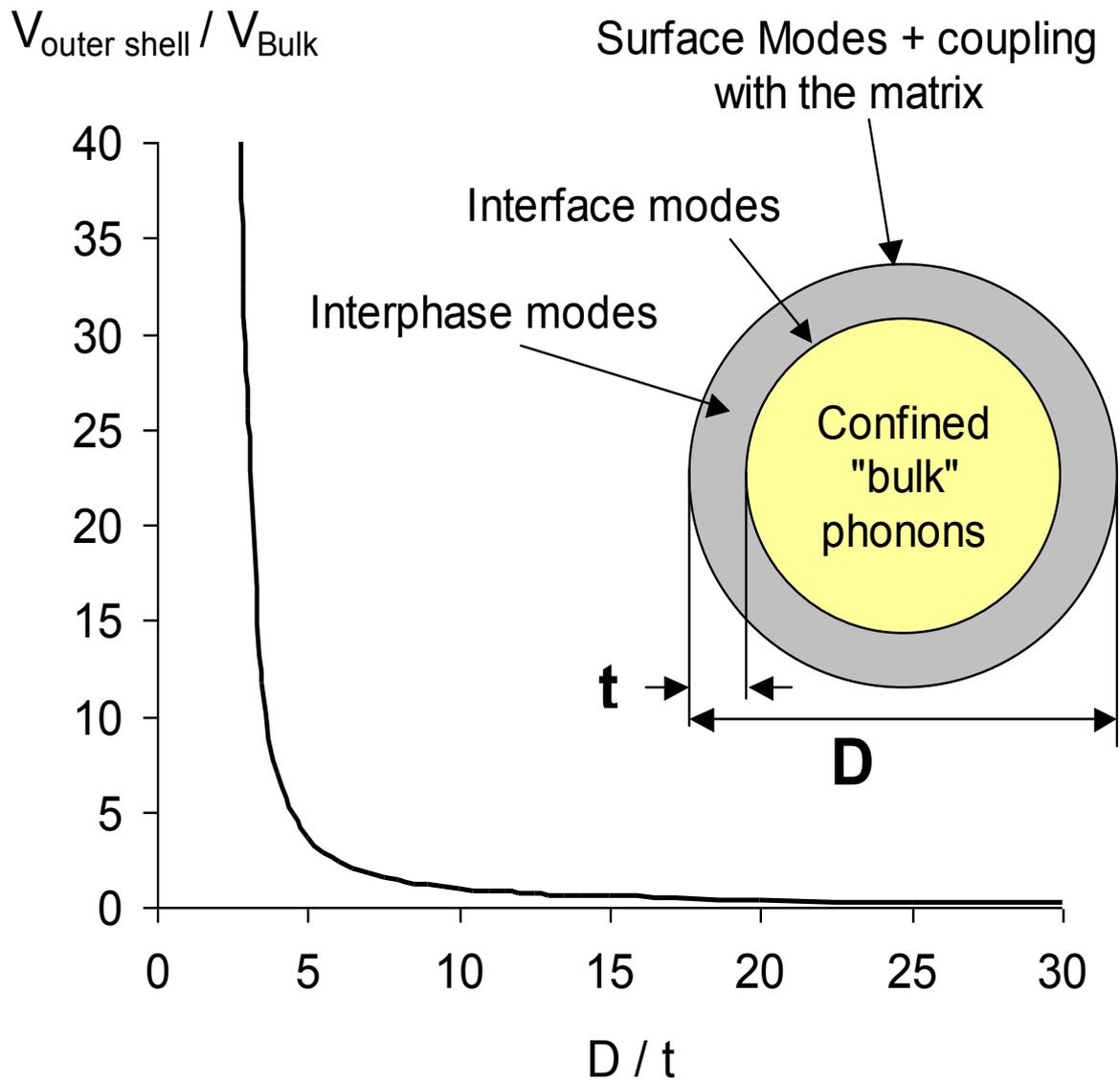

Fig.2

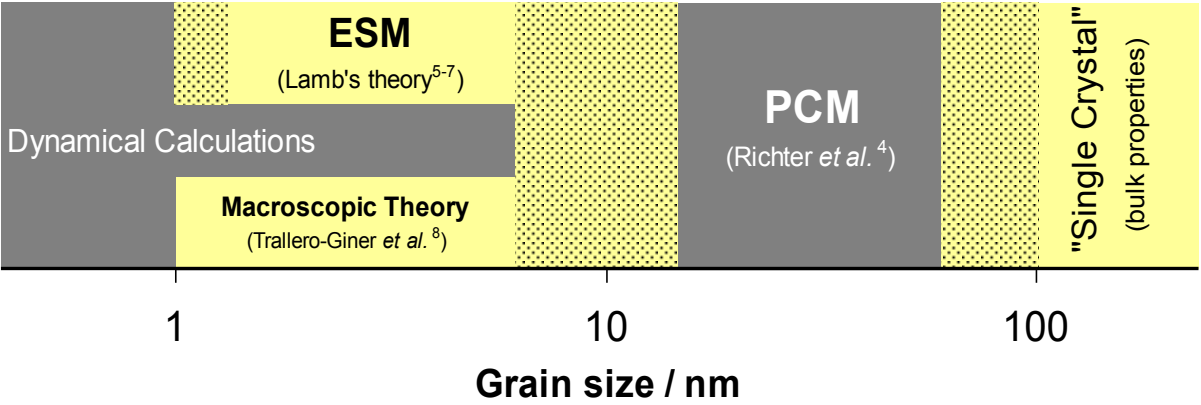